\documentclass[reprint,showpacs,preprintnumbers,amsmath,aps,pre]{revtex4-1}

\begin{document}

\title{Efficiency at maximum power output of linear irreversible Carnot-like heat engines}
\author{Yang Wang}
\affiliation{Department of Physics, Beijing Normal University, Beijing 100875, China}
\author{Z. C. Tu}\email{tuzc@bnu.edu.cn}
\affiliation{Department of Physics, Beijing Normal University, Beijing 100875, China}

\date{\today}

\begin{abstract}
The efficiency at maximum power output of linear irreversible Carnot-like heat engines is investigated based on the assumption that the rate of irreversible entropy production of working substance in each ``isothermal" process is a quadratic form of heat exchange rate between the working substance and the reservoir. It is found that the maximum power output corresponds to minimizing the irreversible entropy production in two ``isothermal" processes of the Carnot-like cycle, and that the efficiency at maximum power output has the form as $\eta_{mP}={\eta_C}/(2-\gamma\eta_C)$ where $\eta_C$ is the Carnot efficiency while $\gamma$ depends on the heat transfer coefficients between the working substance and two reservoirs. The value of $\eta_{mP}$ is bounded between $\eta_{-}\equiv \eta_C/2$ and $\eta_{+}\equiv\eta_C/(2-\eta_C)$. These results are consistent with those obtained by Chen and Yan [J. Chem. Phys. \textbf{90}, 3740 (1989)] based on the endoreversible assumption, those obtained by Esposito \textit{et al.} [Phys. Rev. Lett. \textbf{105}, 150603 (2010)] based on the low-dissipation assumption, and those obtained by Schmiedl and Seifert [EPL \textbf{81}, 20003 (2008)] for stochastic heat engines which in fact also satisfy the low-dissipation assumption. Additionally, we find that the endoreversible assumption happens to hold for Carnot-like heat engines operating at the maximum power output based on our fundamental assumption, and that the Carnot-like heat engines that we focused does not strictly satisfy the low-dissipation assumption, which implies that the low-dissipation assumption or our fundamental assumption is a sufficient but non-necessary condition for the validity of $\eta_{mP}={\eta_C}/(2-\gamma\eta_C)$ as well as the existence of two bounds $\eta_{-}\equiv \eta_C/2$ and $\eta_{+}\equiv\eta_C/(2-\eta_C)$.
\end{abstract}
\maketitle

\section{\label{sec:level1} Introduction}
One of the motivations of the early development of thermodynamics was to optimize heat engines
which transform partially heat into work. In 1824, Carnot arrived at the conclusion that the
efficiency of a heat engine operating between two reservoirs at different temperatures can reach the maximum value only for the reversible (quasi-static) processes and depends only on
the relative temperature of two reservoirs. The quasi-static processes
imply the time for performing a Carnot cycle is infinitely large, while the work done by the heat engine in the cycle is finite. Thus the power output of Carnot heat engines is vanishing. A more meaningful case in reality is to search for engines working at much higher power output and estimate the corresponding efficiency.  To achieve the non-vanishing power output, the Carnot cycle should be speeded up and performed in finite time. This issue is of great importance from the economical perspective in our times of soaring oil price.

The problem of efficiency at maximum power output of heat engines has been attracted much attention \cite{Curzon1975,Salamon80,Chen1989,ChenJC94,Bejan96,ChenL99,vdbrk2005,dcisbj2007,Schmiedl2008,Tu2008,Esposito2009a,Esposito2009,Esposito2010,EspositoPRE10,GaveauPRL10,Izumida,wangx10,Velasco10,Abe2011}.
There are several key theoretical developments among them. To begin with, Curzon and Ahlborn \cite{Curzon1975} considered a heat engine undergoing a Carnot-like cycle which consists of two adiabatic processes and two ``isothermal" processes where the working substance in fact does not remain the same temperature as the reservoirs because the the cycle is operated in a finite time. Based on the endoreversible assumption and the heat transfer law that the heat exchange between the working substance and each reservoir is proportional to the temperature difference between them, Curzon and Ahlborn derived the efficiency of Carnot-like engines at maximum power output as
\begin{equation}
 \eta_{CA}=1-\sqrt{1-\eta_C},
\label{Eq-CA}
\end{equation}
where $\eta_C=1- T_3/T_1$ is the Carnot efficiency. Here $T_1$ and $T_3$ are respectively the temperature of the hot and the cold reservoirs. Furthermore, Chen and Yan \cite{Chen1989} discussed the effect of heat transfer law and derived the efficiency of Carnot-like engines at maximum power output as
\begin{equation}
\eta_{CY}=\frac{\eta_C}{2-\gamma_{CY}\eta_C}
\label{Eq-CY}
\end{equation}
when the heat exchange between the working substance and each reservoir is proportional to the difference of inverse temperatures. In the above equation, $\gamma_{CY}=1/(1+\sqrt{\beta/\alpha})$ where $\alpha$ and $\beta$ represent the heat transfer coefficients between the working substance and the hot or the cold reservoirs, respectively. Additionally,
Schmiedl and Seifert \cite{Schmiedl2008} constructed a stochastic heat engine by using an optic trap to control a Brownian particle to perform a Carnot-like cycle and found the efficiency at maximum power output of that engine to be
\begin{equation}
\eta_{SS}=\frac{\eta_C}{2-\gamma_{SS}\eta_C}
\label{Eq-SS}
\end{equation}
with $\gamma_{SS}=1/(1+\sqrt{A_3/A_1})$ within the framework of stochastic
thermodynamics \cite{Sekimoto97,Seifert02,Schmiedl07}.
$A_1$ and $A_3$ are called the irreversible ``actions" of ``isothermal" processes in contact with the hot and cold reservoirs, respectively. As a byproduct, they also proved that the minimum irreversible entropy production in an ``isothermal" process was inversely proportional to the time required to complete that process. In addition, one of the present authors \cite{Tu2008} investigated the Feynman ratchet and found the efficiency at maximum power output of that engine to be
\begin{equation}
\eta_{T}=\frac{\eta_C^2}{\eta_C-(1-\eta_C)\ln{(1-\eta_C)}}.
\label{eq-etaT}
\end{equation}
He also proposed a conjecture that ``a universal efficiency at maximum power, $\eta_C/2  +  \eta_C^2/8$, should exist at small relative temperature differences" \cite{Tu2008}. The work on the thermoelectric efficiency at maximum power in a quantum dot by Esposito \textit{et al.} made this conjecture more clear \cite{Esposito2009a}.

Recently, two key advances on the universality of efficiency at maximum power were achieved by Esposito and his coworkers. They constructed a quite general Carnot-like engine and verified that the efficiency at maximum power exists universality up to quadratic order for strong coupling system in the presence of a left-right symmetry \cite{Esposito2009}. Additionally, they introduced a Carnot-like engine working in the low-dissipation region that the irreversible entropy production in an ``isothermal" process is inversely proportional to the time required to complete that process (this ansatz is called the low-dissipation assumption in the present paper), and then derived the efficiency at maximum power as \cite{Esposito2010}:
\begin{equation}
\eta_{E}=\frac{\eta_C}{2-\gamma_E\eta_C},\label{EqEKLV}
\end{equation}
where $\gamma_E=1/(1+\sqrt{{T_3\Sigma_c}/{T_1\Sigma_h}})$. Here $\Sigma_h$ and $\Sigma_c$ are the proportional coefficients between irreversible entropy production and inverse time for the ``isothermal" processes in contact with the hot and the cold reservoirs, respectively. This result is bounded between $\eta_{-}\equiv \eta_C/2$ and $\eta_{+}\equiv\eta_C/(2-\eta_C)$ \cite{Esposito2010}. By adopting a quite different way, Gaveau and his coworkers proposed a novel definition of efficiency (the sustainable efficiency) and proved that the sustainable efficiency has the upper bound ${1}/{2}$, based on which they also obtained the upper bound $\eta_+={\eta_C}/(2-\eta_C)$ for the efficiency of Carnot-like engines  at maximum power output \cite{GaveauPRL10}.

It is surprisingly interesting that Eqs.~(\ref{Eq-CY}), (\ref{Eq-SS}) and (\ref{EqEKLV}) have the same form as
\begin{equation}
\eta_{mP}=\frac{\eta_C}{2-\gamma\eta_C}\label{uni-etamP}
\end{equation}
except that the parameter $\gamma$ has different meanings. $\eta_{mP}$ is bounded between $\eta_{-}$ and $\eta_{+}$ because of $0<\gamma<1$ in all cases. The work by Esposito \textit{et al.} implies that the low-dissipation assumption is a sufficient condition for the validity of Eq.~(\ref{uni-etamP}) as well as the existence of two bounds $\eta_{-}$ and $\eta_{+}$. As it is mentioned above, the stochastic heat engine obviously satisfies the low-dissipation assumption, thus it is not surprised that Eq.~\eqref{Eq-SS} has the same form as Eq.~\eqref{uni-etamP}.
A series of natural questions are: Does the Carnot-like engine considered by Chen and Yan satisfy the low-dissipation assumption? Further, is the low-dissipation assumption also the necessary condition for the validity of Eq.~\eqref{uni-etamP} as well as the existence of two bounds $\eta_{-}$ and $\eta_{+}$? In this paper, we mainly focus on answering these questions. By using the linear irreversible thermodynamics and assuming that the irreversible entropy production merely comes from the heat exchange between the reservoir and the working substance in the ``isothermal" process, we prove that the Carnot-like engine considered by Chen and Yan does not strictly satisfy the low-dissipation assumption, and that the low-dissipation assumption is a sufficient but non-necessary condition for the validity of Eq.~\eqref{uni-etamP} as well as the existence of two bounds $\eta_{-}$ and $\eta_{+}$. The rest of this paper is organized as follows. In Sec.~\ref{sec-model}, we briefly introduce our theoretical model (linear irreversible Carnot-like engine) and key assumptions. Unlike the model considered by Curzon and Ahlborn, we discard the endoreversible assumption and the assumption that the working substance in each ``isothermal" process has a constant effective temperature, but replace them with a more fundamental assumption that the rate of irreversible entropy production in an ``isothermal" process is a quadratic form of heat exchange rate between the  working substance and the reservoir (i.e., the linear irreversible thermodynamics holds). In Sec.~\ref{sec-optim}, the power output is optimized and the corresponding efficiency is derived, which has the same form as Eq.~\eqref{uni-etamP}. In Sec.~\ref{sec-rel-endor}, we will prove that our assumption includes the endoreversible assumption. In Sec.~\ref{sec-relation}, we discuss the relation between irreversible entropy production and time, and then check it with the low-dissipation assumption. The last section is a brief summary.

\section{Theoretical Model and Basic assumptions \label{sec-model}}
In this section, our theoretical model---a heat engine performing a Carnot-like cycle in finite time is introduced and then the basic assumptions are listed as the basis of our further discussions.

\subsection{Theoretical model: the Carnot-like cycle}
Our research object is a heat engine performing Carnot-like cycle consisting of the following four processes.

1. \textit{``Isothermal" expansion process}.
In this process, the working substance expands in contact with a hot reservoir at temperature $T_1$ and absorbs heat $Q_1$ from the hot reservoir
during the time interval $0<t<t_1$. In this finite-time process, the effective temperature of working substance might be different from $T_1$ and it can vary with time. This is the reason why we add quotation marks on the word ``isothermal" in the whole paper. The total entropy production in this process is
\begin{equation}
\Delta S_1=\frac{Q_1}{T_1} +\Delta S_1^{ir},
\label{Eq-deltaS1}
\end{equation}
where $\Delta S_1^{ir}$ is the irreversible entropy production which is always nonnegative.

2. \textit{Adiabatic expansion process}.
Adiabatic expansion process is idealized as the working substance suddenly decouples from the hot reservoir and then comes into contact with the cold reservoir instantaneously, or equivalently, the temperature of the reservoir is switched from $T_1$ to $T_3$. Simultaneously, some constraints are released from the working substance, for example, enlarging its volume. Here we have in fact assumed that the time for completing adiabatic processes is much shorter than that for completing ``isothermal" processes. There is no heat exchange in this process, i.e. $Q_2=0$.
The variation of entropy in adiabatic expansion process is also vanishing, i.e. $\Delta S_2=0$.

3. \textit{``Isothermal" compression process}.
In this process, the working substance is compressed in contact with a cold reservoir at temperature $T_3$ and releases heat $Q_3$ to the cold reservoir. The time for completing this process is assumed to be $t_3$. The total entropy production in this process is
\begin{equation}
 \Delta S_3 = - \frac{Q_3}{T_3}+\Delta S_3^{ir},
\label{Eq-deltaS3}
\end{equation}
where $\Delta S_3^{ir}$ is the irreversible entropy production which is also nonnegative.

4. \textit{Adiabatic compression process}.
Similar to the adiabatic expansion process, the temperature of reservoir is suddenly switched from $T_3$ to $T_1$. Simultaneously, some constraints are imposed on the working substance, for example, shrinking its volume.
In this process, both the heat exchange and the variation of entropy are vanishing, i.e. $Q_4=0$ and $\Delta S_4=0$.

\subsection{Basic assumptions and direct consequences \label{sec-assump}}
To continue our analysis, we take the following key assumptions.

(i) After going through the Carnot-like cycle, the working substance comes back to its initial state again. Thus there are no net energy change and entropy production in the whole cycle.

(ii) There is no heat leak between the hot and the cold reservoirs. The irreversible entropy is merely produced in the two 'isothermal' processes due to the heat transfers between the substance and its surrounding reservoirs. The prefactor is the inverse of heat transfer coefficient $\kappa_i~(i=1,3)$. The rate of irreversible entropy production in each ``isothermal" process is a quadratic form of heat exchange rate between the working substance and the reservoir (i.e., the linear irreversible thermodynamics holds).

To make this assumption more specific, the thermodynamic force in the ``isothermal'' expansion process may be expressed as $F_1=\frac{1}{T_{1e}(t)}-\frac{1}{T_1}$, where $T_{1e}(t)$ is the effective temperature of the working substance and it is not necessary to presume this effective temperature as constant. The corresponding flux (the rate of heat exchange) can be expressed as $q_1=\dot{Q_1}=\kappa_1 F_1$ which can be time-dependent. Here $\kappa_1$ is the coefficient of heat transfer in the ``isothermal'' expansion process. The rate of irreversible entropy
production in the ``isothermal'' expansion process is then given as $\sigma_1=F_1q_1={q^2_1}/{\kappa_1}$. Similarly, the rate of irreversible entropy production in the ``isothermal'' compression process can be expressed as $\sigma_3={q^2_3}/{\kappa_3}$ where $q_3$ and $\kappa_3$ are the rate of heat exchange and heat transfer coefficient in the ``isothermal'' compression process, respectively.

The above assumptions will result in some direct consequences (The detailed proofs are shown in Appendix \ref{appx-proof}).

(a) The work output of the heat engine in the Carnot-like cycle is
\begin{equation}\label{workoutput}
W=Q_1-Q_3.
\end{equation}
The total entropy productions in two ``isothermal" processes satisfy
\begin{equation}\label{entropy13}-\Delta S_3=\Delta S_1 \equiv\Delta S.\end{equation}

(b) The minimum irreversible entropy production in two ``isothermal" processes for given $t_1$ and $t_3$ can be expressed as
\begin{equation}\label{MEntP}
\min\{\Delta S_j^{ir} \}= Q_j^2/\kappa_j t_j,~(j=1,3).
\end{equation}

\section{Efficiency at maximum power output\label{sec-optim}}
The power output can be defined as $P=W/(t_1+t_3)$. Considering Eqs.~\eqref{Eq-deltaS1}--\eqref{entropy13}, we obtain
\begin{equation}
P=\frac{(T_1-T_3)\Delta S-(T_1\Delta S_1^{ir} + T_3 \Delta S_3^{ir})}{t_1+t_3}.
\label{Eq-power}
\end{equation}
Because $\Delta S$ is a state variable depending only on the initial and final states of the ``isothermal" process while $\Delta S_1^{ir}$ and $\Delta S_3^{ir}$ are process variables depending on the detailed protocols, maximizing the power output implies minimizing $\Delta S_1^{ir}$ and $\Delta S_3^{ir}$ with respect to the protocols for given $t_1$ and $t_3$ and then maximizing $P$ with respect to $t_1$ and $t_3$. In fact, this argument has been presented in different appearances for stochastic heat engines \cite{Schmiedl2008} or quantum-dot Carnot engines \cite{EspositoPRE10}.

When $\Delta S_1^{ir}$ takes the minimum value for given $t_1$, from Eqs.~\eqref{Eq-deltaS1}, \eqref{entropy13} and \eqref{MEntP} we obtain
\begin{equation}\label{eq-Q1}\Delta S=\frac{Q_1}{T_1}+\frac{Q_1^2}{\kappa_1 t_1},
\end{equation}
from which we obtain the solution
\begin{equation}\label{eq-sQ1}
Q_1=\frac{\kappa_1 t_1}{2 T_1}\left(\sqrt{1+\frac{4\Delta S T_1^2}{\kappa_1 t_1}}-1\right).
\end{equation}
Substituting this equation into Eq.~\eqref{MEntP}, we obtain
\begin{equation}
\min\{\Delta S_1^{ir}\}=\frac{\kappa_1 t_1}{2 T_1^2}\left(1+\frac{2\Delta S T_1^2}{\kappa_1 t_1}-\sqrt{1+\frac{4\Delta S T_1^2}{\kappa_1 t_1}}\right).
\label{Eq-minS1}
\end{equation}

Similarly, when $\Delta S_3^{ir}$ takes the minimum value for given $t_3$, we can obtain
\begin{equation}\label{eq-Q3}-\Delta S=-\frac{Q_3}{T_3}+\frac{Q_3^2}{\kappa_3 t_3},
\end{equation}
from which we arrive at
\begin{equation}\label{eq-sQ3}
Q_3=\frac{\kappa_3 t_3}{2 T_3}\left(1-\sqrt{1-\frac{4\Delta S T_3^2}{\kappa_3 t_3}}\right)
\end{equation}
and
\begin{equation}
\min\{\Delta S_3^{ir}\}=\frac {\kappa_3 t_3}{2 T_3^2}\left(1-\frac{2\Delta S T_3^2}{\kappa_3 t_3}-\sqrt{1-\frac{4\Delta S T_3^2}{\kappa_3 t_3}}\right).
\label{Eq-minS3}
\end{equation}

Substituting Eqs.~\eqref{Eq-minS1} and \eqref{Eq-minS3} into Eq.~\eqref{Eq-power}, we have
\begin{equation}
P=\frac{\frac{\kappa_1 t_1}{2 T_1}\left(\sqrt{1+\frac{4\Delta S T_1^2}{\kappa_1 t_1}}-1\right)-\frac{\kappa_3 t_3}{2 T_3}\left(1-\sqrt{1-\frac{4\Delta S T_3^2}{\kappa_3 t_3}}\right)}
{t_1+t_3}.
\label{Eq-power2}
\end{equation}
Now we maximize $P$ with respect to $t_1$ and $t_3$. Taking $\partial P/\partial t_1$ and $\partial P/\partial t_3$ equal to zero, and then introducing dimensionless parameters $\tau=T_3/T_1$, $\upsilon=\kappa_1/\kappa_3$, $M=2 Q_3/\Delta S T_3={\kappa_3 t_3}/{\Delta S T_3^2}-\sqrt{\left({\kappa_3 t_3}/{\Delta S T_3^2}\right)^2-{4\kappa_3 t_3}/{\Delta S T_3^2}}$, and $N=2Q_1/\Delta S T_1=\sqrt{\left({\kappa_1 t_1}/{\Delta ST_1^2}\right)^2+{4\kappa_1 t_1}/{\Delta ST_1^2}}-{\kappa_1 t_1}/{\Delta ST_1^2}$, we can derive
\begin{eqnarray}
  \frac{(N-2)^2}{N(4-N)} &=& \frac{N-\tau M}{\frac{N^2}{2-N}+\tau^2\upsilon \frac{M^2}{M-2}},\label{eq-NM1}\\
  \frac{(M-2)^2}{M(4-M)} &=& \frac{\tau \upsilon (N-\tau M)}{\frac{N^2}{2-N}+\tau^2\upsilon \frac{M^2}{M-2}}.\label{eq-NM2}
\end{eqnarray}
Solving the above two equations, we arrive at
\begin{equation}\label{solutinN}
N=\frac{4\tau M}{\tau M+4-M}
\end{equation}
and
\begin{equation}\label{solutinM}
M=\frac{4\sqrt{\upsilon}+4}{(\tau+1)\sqrt{\upsilon}+2}.
\end{equation}
The details will be shown in Appendix \ref{appx-deriv}.

Now we calculate the efficiency at maximum power output.
As we know, the efficiency is defined as $\eta\equiv W/Q_1=1-Q_3/Q_1$, which can be expressed as $\eta=1-\tau M/N$
by introducing $N$ and $M$. Considering Eqs.~\eqref{solutinN} and \eqref{solutinM}, we can derive the efficiency at maximum power output of Carnot-like heat engines which has the same form as Eq.~\eqref{uni-etamP} with $\gamma=1/(1+\sqrt{\kappa_3/\kappa_1})$.
This formula has the same form as the result derived by Chen and Yan although we take an assumption different from the endoreversible assumption. The relation between our assumption (ii) and the endoreversible assumption will be further discussed in the following section.

\section{The relation between our assumption and the endoreversible assumption\label{sec-rel-endor}}
As it is shown in Appendix \ref{appx-proof}, when the irreversible entropy production in each ``isothermal" process takes the minimum value, the rate of heat exchange satisfies $q_i= Q_i/t_i~(i=1,3)$ which is independent on time. That is, the effective temperature of the working substance is unchanged in each ``isothermal" processes.
On the other hand, our assumption (ii) implies that the rate of heat exchange and the thermodynamics force display the linear behavior. For the ``isothermal" expansion process, this behavior may be expressed as $q_1 = \kappa_1(1/T_{1e}- 1/T_1)$, where $T_{1e}$ is the effective temperature of the working substance in the ``isothermal" expansion process. Thus we can derive
\begin{equation}
\kappa_1\left(\frac{1}{T_{1e}}- \frac{1}{T_1}\right) = \frac{Q_1}{t_1},
\end{equation}
from which we can further obtain
\begin{equation}
\frac{Q_1}{T_{1e}}= \frac{Q_1}{T_1} + \frac{Q_1^2}{\kappa_1 t_1}.\label{Q1overT1e}
\end{equation}

Similarly, we can derive
\begin{equation}
\frac{Q_3}{T_{3e}}= \frac{Q_3}{T_3} - \frac{Q_3^2}{\kappa_3 t_3},\label{Q3overT3e}
\end{equation}
where $T_{3e}$ is the effective temperature of the working substance in the ``isothermal" compression process. Combining Eqs.~\eqref{eq-Q1}, \eqref{eq-Q3}, \eqref{Q1overT1e} and \eqref{Q3overT3e}, we obtain
\begin{equation}\label{eq-endrevs}
\frac{Q_1}{T_{1e}}-\frac{Q_3}{T_{3e}}=0.
\end{equation}
This formula is no more than the endoreversible assumption in Ref.~\cite{Curzon1975}.
From this analysis, we can see that our assumption implies that the endoreversible assumption happens to hold for Carnot-like heat engines operating at the maximum power output (or minimum the irreversible entropy production in each finite-time ``isothermal" processes). Thus our assumption is more general and fundamental than the endoreversible assumption.

\section{The relation between irreversible entropy production and time\label{sec-relation}}
It is hard to derive the relation between the irreversible entropy production in an ``isothermal" process and the time for completing that process. However, the minimum irreversible entropy production and the time display relative simple relations shown in Eqs.~\eqref{Eq-minS1} and \eqref{Eq-minS3}. Obviously, although we obtain the same form of efficiency at maximum power and its two bounds as Ref.~\cite{Esposito2010}, $\min\{\Delta S_1^{ir}\}$ and $\min\{\Delta S_3^{ir}\}$ are not inversely proportional to the time for completing the ``isothermal" processes. That is, the low-dissipation assumption is a sufficient but non-necessary condition for the validity of Eq.~\eqref{uni-etamP} as well as the existence of two bounds $\eta_{-}$ and $\eta_{+}$.

The relation between our assumption and the low-dissipation assumption \cite{Esposito2010} is discussed as follows. If ${\Delta S T_i^2}/{\kappa_i t_i}\ll 1$, expanding Eqs.~\eqref{Eq-minS1} and \eqref{Eq-minS3} into Taylor series and then keeping the lowest-order terms, we obtain
\begin{equation}
\min\{\Delta S_j^{ir} \} \approx \frac{\Delta S^2 T_j^2}{\kappa_j t_j},~~(j=1,3),
\label{minS-longt}
\end{equation}
which accords with the form of the low-dissapation assumption that the irreversible entropy production is inversely proportional to the time for completing the ``isothermal" processes.
In this case, Eqs.~\eqref {Eq-deltaS1} and \eqref{Eq-deltaS3} can be transformed into
\begin{equation}
 Q_1=T_1(\Delta S-{\Delta S^2 T_1^2}/{\kappa_1 t_1}), \\ \label{Q_1re}
\end {equation}
and
\begin{equation}
 Q_3=T_3(\Delta S+{\Delta S^2 T_3^2}/{\kappa_3 t_3}).\label{Q_3re}
\end{equation}
It is easy to find that the two equations above are in similar form as Eq.~(5) in Ref.~\cite{Esposito2010} if ${\Delta S^2 T_1^2}/{\kappa_1}$ and ${\Delta S^2 T_3^2}/{\kappa_3}$ correspond to $\Sigma_h$ and $\Sigma_c$ in Ref.~\cite{Esposito2010}, respectively. Further, we can also find the similar relation between the two equations above and Eq.~(28) in Ref.~\cite{EspositoPRE10} if ${\Delta S^2 T_1^2}/{\kappa_1}$ and ${\Delta S^2 T_3^2}/{\kappa_3}$ correspond to $(\phi_1-\phi_0)^2/C_h$ and $(\phi_1-\phi_0)^2/C_c$ in Ref.~\cite{EspositoPRE10}, respectively.
Substituting the two equations above into Eq.~\eqref{Eq-power} and then maximizing the power output,
we obtain the optimized time $(t_1^{\ast}  \textrm{ and }  t_3^{\ast})$ for completing two `isothermal' processes which satisfies
\begin{equation}
\frac{\Delta S T_1^2}{\kappa_1 t_1^{\ast}} = (\frac{T_1-T_3}{2T_1})\frac{1}{1+\sqrt{{\kappa_1 T_3^3}/{\kappa_3 T_1^3}}} , \label{optimise1}
\end{equation}
and
\begin{equation}
 \frac{\Delta S T_3^2}{\kappa_3 t_3^{\ast}} = (\frac{T_1-T_3}{2T_3})\frac{1}{1+\sqrt{{\kappa_3 T_1^3}/{\kappa_1 T_3^3}}} .\label{optimise2}
\end{equation}
According to Eqs.~\eqref{optimise1} and \eqref{optimise2}, we can see that ${\Delta S {T_i}^2}/{\kappa_i t_i^{\ast}}\ll 1$ if the temperature difference between the reservoirs is small enough $({T_1-T_3} \ll {T_1})$, which ensures us to derive Eq.~\eqref{minS-longt} from Eqs.~\eqref{Eq-minS1} and \eqref{Eq-minS3}. Thus in this case our linear irreversible engines are degenerated into the low-dissapation engines. The requirement for small temperature difference between the reservoirs is the same as the first condition to guarantee low-dissipation derived by Esposito \emph{et al.} for quantum-dot Carnot engines \cite{EspositoPRE10}. However, we have not found the correspondence to the second condition to ensure low-dissipation for quantum-dot Carnot engines \cite{EspositoPRE10} because our model system based on the linear irreversible thermodynamics is different from theirs \cite{EspositoPRE10}.
The correspondence to the second condition \cite{EspositoPRE10} might be found in nonlinear irreversible Carnot-like heat engines such as the minimally nonlinear irreversible model system \cite{Izumida11}.

Considering Eqs.~\eqref{Q_1re}--\eqref{optimise2}, we can obtain that the efficiency at maximum power output has the same form as Eq.~\eqref{uni-etamP} with $\gamma=1/(1+\sqrt{{\kappa_1 T_3^3}/{\kappa_3 T_1^3}})$ through a few steps of calculations. Here $\gamma$ has the similar form as $\gamma_E=1/(1+\sqrt{{\Sigma_c T_3}/{\Sigma_h T_1}})$ in Ref.~\cite{Esposito2010}, which is consistent with the correspondence relation (${\Delta S^2 T_1^2}/{\kappa_1}$ and ${\Delta S^2 T_3^2}/{\kappa_3}$ correspond to $\Sigma_h$ and $\Sigma_c$ respectively) mentioned above.

It is interesting that the results of low-dissipation engines and our theoretical model give the same bounds $\eta_-\equiv \eta_C/2$ and $\eta_+\equiv \eta_C/(2-\eta_C)$, which seems to imply that the lower and upper bounds might hold for Carnot-like engines in more general conditions than low-dissipation assumption or our assumption (ii).

\section{Conclusion and discussion \label{sec-con}}
The efficiency at maximum power output of Carnot-like heat engines and its bounds can be obtained by using the conventional endoreversible assumption or the low-dissipation assumption.
In this paper, we have presented a new viewpoint to estimate this efficiency. It is found that the maximum power output corresponds to minimizing the irreversible entropy production in two ``isothermal" processes in the Carnot-like cycle. Our discussions are mainly based on the assumption that the rate of irreversible entropy production in an ``isothermal" process is a quadratic form of heat exchange rate between the working substance and the reservoir. Although this assumption is different from the endoreversible assumption or low-dissipation assumption, we still obtain that the efficiency at maximum power output of Carnot-like heat engines has the same form as Eq.~\eqref{uni-etamP} which is bounded between $\eta_{-}\equiv \eta_C/2$ and $\eta_{+}\equiv\eta_C/(2-\eta_C)$. Therefore, the low-dissipation assumption is a sufficient but non-necessary condition for the validity of Eq.~\eqref{uni-etamP} as well as the existence of two bounds $\eta_{-}$ and $\eta_{+}$.

On the other hand, if the temperature difference between heat reservoirs is small enough,  we found that the minimum irreversible entropy production in an ``isothermal" process is inversely proportional to the time that the process is performed. That is, the low-dissipation assumption is indeed valid for small enough temperature difference between heat reservoirs for our linear irreversible engines. Thus it is not surprised that the same bounds as the low-dissipation engines can be derived in this case. However, we also find that the results with large temperaure difference between the heat reservoirs (in this case our theoretical model is no longer low-dissipative) have the same bounds $\eta_-$ and $\eta_+$, which suggests that these bounds might hold for Carnot-like engines in more general conditions. Additionally, the efficiencies at maximum power output of Feynman ratchet and other kinds of engines are also located in the region bounded between $\eta_-$ and $\eta_+$ \cite{Esposito2010}, which hints that the $\eta_{-}\equiv \eta_C/2$ and $\eta_{+}\equiv\eta_C/(2-\eta_C)$ might hold for the majority heat engines working between two reservoirs at different temperature. It is still an open question to prove this point in a general theoretical framework.

\section*{Acknowledgement}
The authors are grateful to the financial supports from National Natural Science Foundation of China (NO.11075015) and the Fundamental Research Funds for the Central Universities. They are also very grateful to Massimiliano Esposito for his kind communications.

\appendix
\section{Proofs of consequences (a) and (b) in Sec.~\ref{sec-assump}\label{appx-proof}}
Assumption (i) implies that the energy change and entropy production in the whole cycle satisfy $\sum_{i=1}^{4}\Delta E_{i}=0$ and $\sum_{i=1}^{4}\Delta S_{i}=0$, respectively. Considering the first law of thermodynamics, we can obtain Eq.~\eqref{workoutput}. From $\Delta S_2=\Delta S_4=0$ and $\sum_{i=1}^{4}\Delta S_{i}=0$, we derive $-\Delta S_3=\Delta S_1$.

From Assumption (ii), we have
\begin{equation}\Delta S_1^{ir}=\int_0^{t_1}\frac{[q_1(t)]^2}{\kappa_1}dt,
\label{Eq-deltS1a}
\end{equation}
where $q_1(t)$ and $\kappa_1$ are the rate of heat exchange and the coefficient of heat transfer in the ``isothermal" expansion process, respectively. Now we will minimize $\Delta S_1^{ir}$ with the constraint
\begin{equation}\label{Eq-constr}
\int_0^{t_1}q_1(t) dt =Q_1\end{equation}
for given $t_1$. Introducing a Lagrange multiplier $\lambda_1$ and then minimizing the unconstrained functional
\begin{equation}I_1 \{q_1\}=\int_0^{t_1}\frac{[q_1(t)]^2}{\kappa_1}dt -\lambda_1\left(\int_0^{t_1}q_1(t)dt-Q_1\right),
 \label{Eq-func1}
\end{equation}
we have $q_1(t)=\kappa_1\lambda_1/2$ which is independent on time. Substituting it into Eqs.~\eqref{Eq-constr} and \eqref{Eq-deltS1a}, we obtain $\min\{\Delta S_1^{ir} \}=Q_1^2/\kappa_1 t_1$.

Similarly, minimizing $\Delta S_3^{ir}=\int_{t_1}^{t_1+t_3}\frac{[q_3(t)]^2}{\kappa_3}dt$ with constraint $\int_{t_1}^{t_1+t_3} q_3(t)dt =Q_3$, we can obtain $\min\{\Delta S_3^{ir} \}=Q_3^2/\kappa_3 t_3$. Thus we arrive at consequence (b).

\section{Deriving and solving Eqs.~\eqref{eq-NM1} and \eqref{eq-NM2} in Sec.~\ref{sec-optim}\label{appx-deriv}}
The maximum power output is found by setting the derivatives of $P$ in Eq.~\eqref{Eq-power2} with respect to $t_1$ and $t_3$ equal to zero. Equation \eqref{Eq-power2} can be
simplified by introducing dimensionless parameters $x={\kappa_1 t_1}/{\Delta ST_1^2}$ and $y={\kappa_3 t_3}/{\Delta S T_3^2}$. The maximum power output is then corresponding to setting
the derivatives of $P$ with respect to $x$ and $y$ equal to zero, which gives\begin{widetext}
\begin{eqnarray}
  \frac{\kappa_1}{T_1}(\frac{x+2}{\sqrt{x^2+4x}}-1)&=&\frac{T_1(\sqrt{x^2+4x}-x)-T_3(y-\sqrt{y^2-4y})}
{{T_1^2 x}/{\kappa_1}+{T_3^2y}/{\kappa_3}}  \\
  \frac{\kappa_3}{T_3}(\frac{y-2}{\sqrt{y^2-4y}}-1)&=& \frac{T_1(\sqrt{x^2+4x}-x)-T_3(y-\sqrt{y^2-4y})}
{{T_1^2x}/{\kappa_1}+{T_3^2y}/{\kappa_3}}
\end{eqnarray}
\end{widetext}
By introducing $\tau=T_3/T_1$, $\upsilon=\kappa_1/\kappa_3$, $M=y-\sqrt{y^2-4y}$ and $N=\sqrt{x^2+4x}-x$, the above two equations are transformed into Eqs.~\eqref{eq-NM1} and \eqref{eq-NM2}. When Eq.~\eqref{eq-NM1} is divided by Eq.~\eqref{eq-NM2}, we derive
\begin{equation}
  \frac{(M-2)^2}{M(4-M)} = \tau \upsilon \frac{(N-2)^2}{N(4-N)}.\label{eq-NM3}
\end{equation}
From the above equation, we can express $M^2/(M-2)$ with other terms. Then substituting it into Eq.~\eqref{eq-NM1}, we arrive at Eq.~\eqref{solutinN}. Next, substituting Eq.~\eqref{solutinN} into Eq.~\eqref{eq-NM3}, we derive
\begin{equation}
 [4-(\tau+1)M]^2 \upsilon=4(M-2)^2.
\label{Eq-MMMM}
\end{equation}
Form $M=y-\sqrt{y^2-4y}$ and $N=\sqrt{x^2+4x}-x$ we know $0<N<2$ and $2<M<4$. Considering Eq.~\eqref{solutinN}, we derive $2<M< 4/(1+\tau)$. Thus Eq.~\eqref{Eq-MMMM} reduces to Eq.~\eqref{solutinM}.

\end{document}